 \documentclass[final,3p,times,twocolumn]{elsarticle}
 \newcommand{\vek}[1]{\mathbf{#1}} %\underline
\newcommand{\nn}[0]{\nonumber}

\newcommand{\mat}[1]{\underline{\underline{\mathbf{#1}}}}

\newcommand{\bfnabla}[0]{\boldsymbol{\nabla}}
\newcommand{\bftaumat}[0]{\mat{\boldsymbol{\tau}}}
\usepackage{graphicx}% Include figure files
\usepackage{dcolumn}% Align table columns on decimal point
\usepackage{bm}% bold math
\usepackage{amsbsy}
\usepackage{multirow}
\usepackage{bbding}
\usepackage[abs]{overpic}
\usepackage[dvips]{color}

\usepackage{amssymb}
 \biboptions{comma,square,numbers,sort}

\journal{arXiv}

\begin{document}

\begin{frontmatter}



\title{A Viscoelastic Catastrophe}


\author{Kristian Ejlebjerg Jensen\corref{cor1}}
\ead{kristian.jensen@nanotech.dtu.dk}
\address{Department of Micro- and Nanotechnology, Technical University of Denmark, DK-2800 Kgs. Lyngby, Denmark}

\author{Peter Szabo}
\ead{ps@kt.dtu.dk}
\address{Department of Chemical and Biochemical Engineering, Technical University of Denmark, DK-2800 Kgs. Lyngby, Denmark}

\author{Fridolin Okkels}%\corref{cor2}}
\ead{fridolin.okkels@fluidan.com}
\address{Diplomvej, Fluidan, DK-2800 Kgs. Lyngby, Denmark}

\cortext[cor1]{Corresponding author}
%\cortext[cor2]{Principal corresponding author}
\begin{abstract}
We use a differential constitutive equation to model the flow of a viscoelastic flow in a cross-slot geometry, which is known to exhibit bistability above a critical flow rate. The novelty lies in two asymmetric modifications to the geometry, which causes a change in the bifurcation diagram such that one of the stable solutions becomes disconnected from the solution at low flow speeds. First we show that it is possible to mirror one of the modifications such that the system can be forced to the disconnected solution. Then we show that a slow decrease of the flow rate, can cause the system to go through a drastic change on a short time scale, also known as a catastrophe. The short time scale could lead to a precise and simple experimental measurement of the flow conditions at which the viscoelastic catastrophe occurs. Since the phenomena is intrinsically related to the extensional rheology of the fluid, we propose to exploit the phenomena for in-line extensional rheometry.
\end{abstract}


\end{frontmatter}

% \linenumbers

\section{\label{sec:intro}Introduction}
\begin{figure}[!htb]
\centering \includegraphics[width= 0.47 \textwidth]{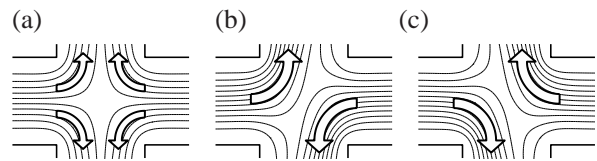}
%\vspace{2mm}
%\begin{minipage}{0.14\textwidth}
%\centering
%\begin{overpic}[tics=10,width=\textwidth]{fig/bif/flowlines/flowb}\put(0,48){(a)}\end{overpic}
%\end{minipage}\hspace{0.014\textwidth}
%\begin{minipage}{0.14\textwidth}
%\centering
%\begin{overpic}[tics=10,width=\textwidth]{fig/bif/flowlines/flowa}\put(0,48){(b)}\end{overpic}
%\end{minipage}\hspace{0.014\textwidth}
%\begin{minipage}{0.14\textwidth}
%\centering
%\reflectbox{\begin{overpic}[tics=10,width=\textwidth]{fig/bif/flowlines/flowa}\put(61,48){\reflectbox{(c)}}\end{overpic}}
%\end{minipage}
\caption{The viscoelastic flow in a cross is symmetric in the regime of low elasticity (a), but the symmetric flow becomes unstable in the regime of moderate elasticity. Instead two asymmetric solutions appear, (b) and (c).}\label{fig:intro}
\end{figure}

Whenever flexible objects or large molecules are dissolved in a fluid, the flow can give rise to stretching and orientation of these constituents, which in turn can lead to stresses that affect the flow such that a feedback is created. This can result in upstream vortices, instabilities or bistability in what is collectively referred to as viscoelastic effects. Experiments with weakly elastic fluids tend to be dominated by inertia, but the relative magnitude of inertial to viscoelastic effects vanishes as the characteristic length scale is reduced, and this scaling law constitutes the foundation for micro-rheometry \cite{pipe2009microfluidic,galindo2013microdevices}. The properties of viscoelastic fluids in shear flow is important for pipe flow and lubrication, but complex flows involving contractions and/or obstacles tend to emphasize the extensional properties of these fluids. The cross-slot geometry illustrated in figure \ref{fig:intro} give rise to an extensional flow near the stagnation point in the center. Therefore it has been suggested as an extensional micro-rheometer, but often in the context of a birefringence setup \cite{haward2012optimized} and with the tendency for bistability at moderate elasticity viewed as a limiting factor. 

The paper is based on simulation of differential constitutive models, and with these tools we predict that it is possible to generate a catastrophe due to viscoelastic effects, that is, an abrupt change in the flow pattern under slow and smooth variations in the input parameters. The abrupt nature of the catastrophe ought to permit detection by simple means, e.g. polarized light \cite{coventry2008cross} or ultrasonic response \cite{takeda2012ultrasonic}, and the conditions at which the catastrophe occurs depends on the extensional rheology of the fluid. Therefore we suggest to use it as an extensional micro-rheometer, as the reproducibility is potentially only limited by the continuity of the fluid and the quality of the pumps. The non-ideal extensional flow is a disadvantage, but many applications are far from ideal flows and in some cases more weight might be put on the simplicity of construction and/or the in-line monitoring capability provided by the confined nature of the device.

The article is structured in four parts: First the modeling of a viscoelastic FENE-CR fluid is described. Then the features of the symmetric cross-slot geometry are sketched, before the changes needed to generate the catastrophe are described on an abstract level. Finally we show a concrete example on how to generate the catastrophe with the two inlet flow rates as parameters for the an asymmetric version of the cross-slot geometry.

\section{Modeling}
The elastic behavior of viscoelastic flow is often modeled using a statistical ensemble of flexible dumbbells (that is two point masses connected by a spring) in what is collectively referred to as dumbbell models. The end-to-end vector, $\vek{a}$ of such a dumbbell describes orientation as well as extension, and it defines the conformation tensor,
\begin{eqnarray}
\mat{A}=\left \langle \vek{a}\otimes\vek{a}\right \rangle /\mathrm{a}_\mathrm{eq}^2 \label{eqn:conf},
\end{eqnarray}
where $\langle \cdots \rangle$ is a statistical average, and $\mathrm{a}_\mathrm{eq}$ is the equilibrium length of the end-to-end vector. 

Balance of stress is guaranteed by the Stokes equation 
\begin{eqnarray}
0 &=& \vek{\bfnabla} \cdot \left(-p\mat{I}+\eta_s \left[\bfnabla \, \vek{v} + \left(\bfnabla \, \vek{v} \right)^T\right] + \bftaumat_e\right) \label{eqn:stokes} \\%- \alpha(\rho) \vek{v} %\rho\frac{D\vek{v}}{Dt} &=& \vek{\bfnabla} \cdot \left(-p\mat{I}+\eta_s \bfsymbmat{\dot{\gamma}} + \bfsymbmat{\tau}_e\right) - \alpha\vek{v} \label{eqn:stokes} \\%- \alpha(\rho) \vek{v} 
0 &=& \vek{\bfnabla} \cdot \vek{v}, \label{eqn:cont}
\end{eqnarray}
where $\vek{v}$ is the velocity, $p$ is the pressure, $\mat{I}$ is the identity matrix and $\eta_s$ is the solvent viscosity. %The damping term, $-\alpha\vek{v}$ is used by the optimization. It vanishes in fluid regions, while solid regions are characterized by a large inverse permeability, $\alpha_\mathrm{max}$ similar to a sponge with low permeability. 
The polymer stress tensor $\bftaumat_e$ describes the stress due to the elastic dumbbells,
\begin{eqnarray}
\bftaumat_e &=& \frac{\eta_p}{\lambda}k(\mat{A})\left(\mat{A}-\mat{I} \right) \label{eqn:tau} \\
k(\mat{A}) &=& \frac{1}{1-\mathrm{Trace}(\mat{A})/\mathrm{a}_\mathrm{max}^2} \label{eqn:k},
\end{eqnarray}
where $\eta_p$ is the polymer viscosity, $\lambda$ is the dumbbell relaxation time and $k(\mat{A})$ is a non-linear modification of the dumbbell spring constant. This modification constrains the maximum extension, by setting $\mathrm{a}_\mathrm{max}^2$ as an upper limit on the trace of the conformation tensor. Finally there is the evolution equation for the conformation tensor, which involve convection on the left-hand side and relaxation together with orientation and stretching on the right-hand side,
\begin{eqnarray}
\frac{D\mat{A}}{Dt} &=&  -\frac{k(\mat{A})}{\lambda} \left(\mat{A}-\mat{I} \right) \nn \\
&+&\left[\mat{A} \cdot \vek{\bfnabla} \, \vek{v} + \left(\vek{\bfnabla} \, \vek{v}\right)^T \cdot \mat{A}\right] , \label{eqn:hyp}
\end{eqnarray}
where $D/Dt$ is the material derivative. Note how the relaxation term will tend to dominate the equation as the trace of the conformation tensor approaches $\mathrm{a}_\mathrm{max}^2$. Equations (\ref{eqn:stokes}-\ref{eqn:hyp}) constitute the FENE-CR model \cite{chilcott1988creeping}, which has a constant shear viscosity. This is contrary to the FENE-P model \cite{bird1980polymer}, which exhibit shear thinning behavior and inspired the FENE-CR model.

Besides $\mathrm{a}_\mathrm{max}$, the equation system only has two dimensionless parameters 
\begin{eqnarray}
\beta = \frac{\eta_s}{\eta_s+\eta_p} \quad &\mathrm{and}& \quad \mathrm{We} = \lambda\frac{v_\mathrm{char}}{L_\mathrm{char}} ,\nn
\end{eqnarray}
where $v_\mathrm{char}$ is a characteristic velocity. $\beta$ is the solvent to total viscosity ratio, and it expresses the proportion of viscous effects due to the solvent. $\beta=1$ thus corresponds to a Newtonian fluid, while $\beta=0$ and $\mathrm{a}_\mathrm{max}\rightarrow\infty$ results in particularly strong elastic effects in what is usually referred to as the Upper Convected Maxwell model. The Weissenberg number, We, expresses the strength of elastic to viscous effects for the dumbbells. %As opposed to the Reynolds number, the Weissenberg number is independent of length scale and it is this basic scaling difference, which constitutes the foundation of micro rheometry.

We solve the set of equations (\ref{eqn:stokes}-\ref{eqn:hyp}) using an implementation \cite{jensen2015implementation} in COMSOL Multiphysics, a commercial high-level finite element package.

\section{The Symmetric Cross-Slot}
\begin{figure}[!htb]
\centering \includegraphics[angle=-90,width= 0.35 \textwidth]{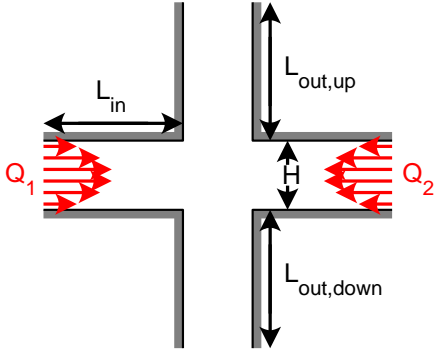}
%\centering \includegraphics[angle=-90,width= 0.35 \textwidth]{fig/geom/geom}
\caption{The cross-slot geometry is illustrated with inlets to the sides and outlets at the top and bottom.} \label{fig:geom}
\end{figure}

For $Q_1=Q_2$ and $L_\mathrm{out,up}=L_\mathrm{out,down}$ figure \ref{fig:geom} illustrate the symmetric cross. We will not show new simulations for this as numerous results already exists \cite{poole2007purely,rocha2009extensibility,jensen2015implementation,jensen2014optimization}, but just describe the qualitative features of the symmetric case.

\begin{figure}[!htb]
\centering \includegraphics[width= 0.47 \textwidth]{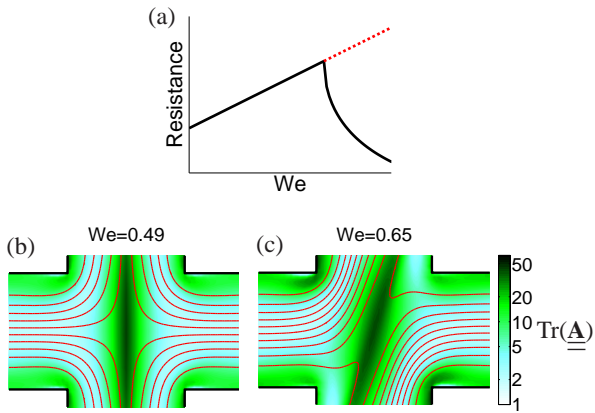}
%\centering \begin{overpic}[tics=10,width=0.2\textwidth]{fig/bif/power}\put(-5,70){(a)}\end{overpic}\\\vspace{3mm}
%%\centering \begin{overpic}[tics=10,width=0.35\textwidth]{fig/sym/disp}\put(0,127){(a)}\end{overpic}\vspace{3mm}
%%\centering \includegraphics[width= 0.45 \textwidth]{intro}
%\begin{minipage}{0.2\textwidth}
%\centering
%\begin{overpic}[trim=0 0 3cm 0,clip=true,tics=10,width=\textwidth]{fig/sym/flow75}\put(0,65){(b)}\end{overpic}
%\end{minipage}\hspace{0.01\textwidth}
%\begin{minipage}{0.24\textwidth}
%\centering
%\begin{overpic}[tics=10,width=\textwidth]{fig/sym/flow50}\put(0,65){(c)}\put(113,30){$\mathrm{Tr}(\mat{A})$}\end{overpic}
%\end{minipage}\hspace{0.01\textwidth}
\caption{(a) is a sketch of the hydraulic in the cross-slot as a function of the Weissenberg number with the maximum corresponding to the point of bistability. The red line represents an unstable symmetric solution. (b) and (c) show the trace of the conformation tensor on a logarithmic scale together with streamlines before and after the bistability sets in. The stream function is computed and used to plot streamlines in red.}\label{fig:sym}
\end{figure}

We define the Weissenberg number using the total flow rate $Q_1+Q_2$,
\begin{eqnarray}
\mathrm{We} = \lambda\frac{Q_1+Q_2}{2H^2} . \label{eqn:We}
\end{eqnarray}
Figure \ref{fig:sym}(a) sketches the hydraulic resistance of the symmetric cross-slot as a function of the Weissenberg number. The point of maximum resistance corresponds to the point of bistability, and characteristic plots of the trace of the conformation tensor are shown in figure \ref{fig:sym}(b) and (c). Note that a similar plot of the dissipation would also give a maximum at the point of bistability in the case of a flow rate driven setup, where as the point of bistability would be a dissipation minimum in the case of a pressure driven setup. Regardless of the setup, the two asymmetric solutions give rise to the same resistance/dissipation for the same Weissenberg number, and one can think of the rotation of the flow as a way for the fluid to avoid strong extension in the central stagnation point.

\section{Generating the Catastrophe\label{sec:cat}}
\begin{figure}[!htb]
\centering \includegraphics[width= 0.47 \textwidth]{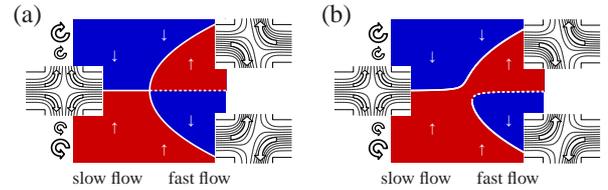}
%\begin{minipage}{0.22\textwidth}
%\centering
%\begin{overpic}[tics=10,width=\textwidth]{fig/bif/perfect_collected2}\put(18,-8){\scriptsize slow flow}\put(55,-8){\scriptsize fast flow}\put(-5,55){(a)}\end{overpic}
%\end{minipage}\hspace{0.03\textwidth}
%\begin{minipage}{0.22\textwidth}
%\centering
%\begin{overpic}[tics=10,width=\textwidth]{fig/bif/aperfect_collected2}\put(18,-8){\scriptsize slow flow}\put(55,-8){\scriptsize fast flow}\put(-5,55){(b)}\end{overpic}
%\end{minipage}\vspace{2mm}
\caption{A pitchfork bifurcation is illustrated in (a) with rotation as solution variable on the y-axis. Red and blue areas represent areas with increasing clock- and counterclockwise rotation, respectively. The insets show the symmetric and two asymmetric solutions at their respective positions in the diagram. Two asymmetric modifications can give rise to the bifurcation shown in (b), where one of the stable solutions has become disconnected from the solution at low flow rates.}\label{fig:intro2}
\end{figure}

We suggest to embrace the bistable character of the cross-slot geometry by changing the nature of this phenomenon such that the point of bistability can be used as a fingerprint of the fluids extensional properties. It is possible to create an abrupt transient effect -- a catastrophe -- by introducing two asymmetries such that the double symmetry of the cross-slot is broken, leaving one of the stable solutions disconnected from the solution at low flow rates, see figure \ref{fig:intro2}(b). 

\begin{figure}[!htb]
\centering \includegraphics[width= 0.47 \textwidth]{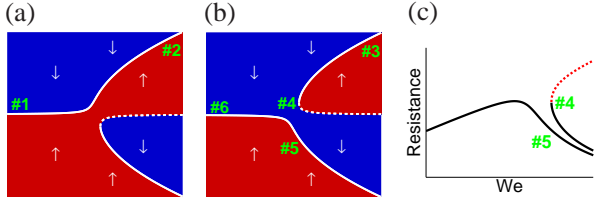}
%\vspace{2mm}
%\begin{minipage}{0.14\textwidth}
%\centering
%\begin{overpic}[tics=10,width=\textwidth]{fig/bif2/bif1}\put(0,65){(a)}\end{overpic}
%\end{minipage}\hspace{0.01\textwidth}
%\begin{minipage}{0.14\textwidth}
%\centering
%\begin{overpic}[tics=10,width=\textwidth]{fig/bif2/bif2}\put(0,65){(b)}\end{overpic}
%\end{minipage}\hspace{0.01\textwidth}
%\begin{minipage}{0.15\textwidth}
%\centering \vspace{2mm}
%\begin{overpic}[tics=10,width=\textwidth]{fig/bif2/apower}\put(2,65){(c)}\end{overpic}
%\end{minipage}
\caption{(a) shows how the system is set in one of the states (\#1 to \#2), while the asymmetry is flipped in (b) (\#3) such that a catastrophe occurs, when the bifurcation parameter is decrease (\#4 to \#5). The catastrophe is associated with a drop in hydraulic resistance as illustrated in (c).}\label{fig:intro3}
\end{figure}

The asymmetry can be introduced in terms of boundary conditions, geometry or perhaps even volume forces, but for our purposes it is imperative that one of the asymmetries can be flipped, as this enables the device to be set in the disconnected state as illustrated in figure \ref{fig:intro3}.  Please take note of the green numbers in figure \ref{fig:intro3}, since we refer to them throughout the rest of the text. $\#1\rightarrow\#2$, the system is set in one of the states, and then one of the asymmetries is flipped [$\#2\rightarrow\#3$], before the general flow rate (the bifurcation parameter) is slowly decreased [$\#3\rightarrow\#6$] which gives rise to a equivalent slow change in the flow until the catastrophe occurs [$\#4\rightarrow\#5$]. This happens at a particular critical flow rate, which is characteristic for the extensional rheology of the working fluid.

\section{The Asymmetric Cross-Slot}
\begin{figure}[!htb]
\centering \includegraphics[angle=-90,width= 0.35 \textwidth]{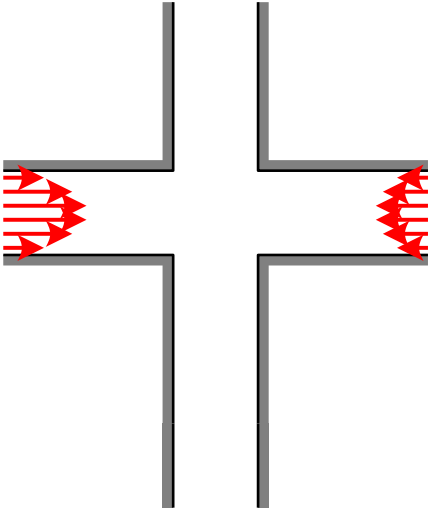}
%\centering \includegraphics[angle=-90,width= 0.35 \textwidth]{fig/geom/geom_asym}
\caption{The asymmetric cross-slot geometry is sketched with different inlet flow rates as well outlet lengths.} \label{fig:ageom}
\end{figure}

The asymmetric cross differs from the symmetric one by having different inlet flow rates as well as outlet lengths as sketched in figure \ref{fig:ageom}. 

\begin{figure}[!htb]
\centering \includegraphics[width= 0.25 \textwidth]{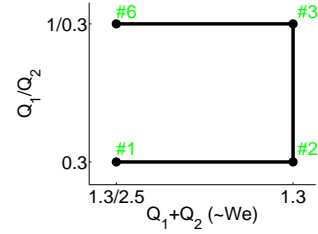}
%\centering \includegraphics[angle=-90,width= 0.35 \textwidth]{fig/geom/geom_asym}
\caption{The flow rates $Q_1$ and $Q_2$ are varies slowly in time to produce the conditions described in section \ref{sec:cat}.} \label{fig:Qs}
\end{figure}

We investigate the case of having an upper outlet length of 9$H$, and a lower outlet length equal to half that. We use the inlet flow rates as control parameters, and we vary them slowly in time to produce the conditions described in section \ref{sec:cat} and illustrated in figure \ref{fig:Qs},
\begin{eqnarray}
Q_1 &=& \left(\xi + \left[1-\xi\right] \mathrm{st}^{\#2}_{\#3}(\tilde{t})\right) \nn \\
&\times& \left(\chi+\mathrm{st}^{\#1}_{\#2}(\tilde{t})\left[1-\chi\right] + \mathrm{st}^{\#3}_{\#6}(\tilde{t})\left[\chi-1\right]\right) \nn \\
Q_2 &=& \left(1 + \left[\xi-1\right] \mathrm{st}^{\#2}_{\#3}(\tilde{t})\right) \nn \\
&\times& \left(\chi+\mathrm{st}^{\#1}_{\#2}(\tilde{t})\left[1-\chi\right] + \mathrm{st}^{\#3}_{\#6}(\tilde{t})\left[\chi-1\right]\right) , \nn
\end{eqnarray}
where $\xi$ is the ratio between the inlet flow rates, while $\chi$ is the ratio of maximum and minimum Weissenberg numbers -- due to the fact that these are proportional to the total flow rate as defined in equation (\ref{eqn:We}). 

We set the maximum Weissenberg number to 1.14, $\xi=0.3$, $\chi=2.5$, $\mathrm{a}_\mathrm{max}^2=100$ and $\beta=0.05$. $\mathrm{st}^{\#1}_{\#2}(\tilde{t})$, $\mathrm{st}^{\#2}_{\#3}(\tilde{t})$ and $\mathrm{st}^{\#3}_{\#6}(\tilde{t})$ are step functions given by 
\begin{eqnarray}
%\mathrm{We} &=& \mathrm{We}_\mathrm{start} + \left(\mathrm{We}_\mathrm{end} -\mathrm{We}_\mathrm{start} \right)\mathrm{st}(\tilde{t}), \quad \mathrm{where} \label{eqn:We} \\
\mathrm{st}(\tilde{t}) &=& \left \{\begin{array}{l l l} 0 &,& \tilde{t} < \tilde{t}_\mathrm{start} \\
0.5+1.5\bar{t}-2\left(\bar{t}\,\right)^3 &, \,\tilde{t}_\mathrm{start} \leq & \tilde{t} < \tilde{t}_\mathrm{end} \\
1 &, \, \tilde{t}_\mathrm{end} \leq & \tilde{t} \end{array}\right. , \label{eqn:step} \\%f(t)=0.5+1.5t-2t^3
\bar{t} &=& \frac{\tilde{t}-(\tilde{t}_\mathrm{start}+\tilde{t}_\mathrm{end})/2}{\tilde{t}_\mathrm{end}-\tilde{t}_\mathrm{start}}\nn \\
T_\mathrm{step}&=&(\tilde{t}_\mathrm{end}-\tilde{t}_\mathrm{start})/\mathrm{We}_\mathrm{max} \nn
\end{eqnarray}
with $\tilde{t}_\mathrm{start}$ equal to 0, $T_\mathrm{step}$ and $2T_\mathrm{step}$, respectively. This way we make a single transient simulation that is in fact a collection of quasi-steady solutions to a range of Weissenberg numbers. For the time stepping we use an implicit scheme with adaptive step size, such that slow variation of the Weissenberg number does not increase the total computation time \cite{multiphysics2008user}. 

\begin{figure}[!htb]
\centering \includegraphics[width= 0.47 \textwidth]{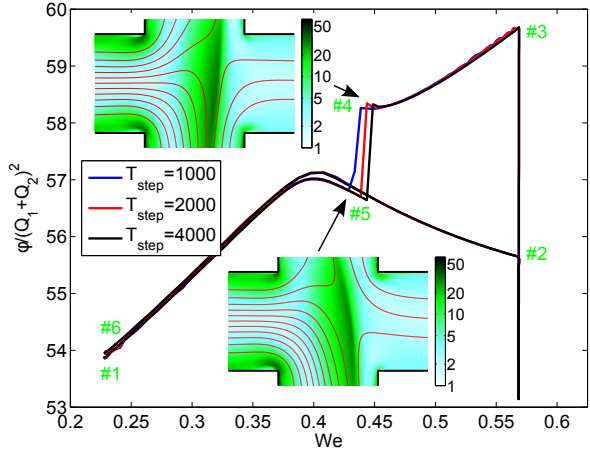}
%\centering \begin{overpic}[tics=10,width=0.45\textwidth]{fig/sim/newq/asym_collected}\end{overpic}
\caption{The dissipation in the asymmetric cross-slot is shown as a function of Weissenberg number for a quasi-steady calculation with insets showing the trace of the conformation tensor on a logarithmic scale together with streamlines in red. The numbers in green corresponds to those in figure \ref{fig:intro3} with the catastrophe between \#4 and \#5. We show results for three different speeds of these transient simulations ($T_\mathrm{step}$) to verify, that the time scale of the jump is independent of this. The simulation is performed with 263k degrees of freedom.} \label{fig:asym}
\end{figure}

We calculate the dissipation (using the Frobenius product denoted with colon)
\begin{eqnarray}
\phi = \int_\Omega \bfnabla \,\vek{v} : \left(\eta_s \left[\bfnabla \, \vek{v} + \left(\bfnabla \, \vek{v} \right)^T\right] + \bftaumat_e\right) d\Omega, \nn
\end{eqnarray}
normalize it with the squared total flow rate, and plot this versus the Weissenberg number in figure \ref{fig:asym} together with insets of the conformation tensor and streamlines before and after the catastrophe. We show results for three different values of $T_\mathrm{step}$, and the fact that the jump appears sharper for the two larger values indicate that the timescale of the jump does not scale with $T_\mathrm{step}$. This is to be expected, since the time scale of the jump ought to be close to the relaxation time. The magnitude of the jump is important for a practical realization, but this is completely dependent on the nature of the detection principle. In example, we estimate the shift in angle of the birefringent strand to 13 degrees in our simulations, which should easily be detectable by optical methods. 

In terms of experimental realization, we would suggest keeping the ratio of the outlet lengths below 2 and also to join them on the micro scale. Together with alertness towards unsteady flows, this should give optimal conditions for catastrophes occurring at reproducible conditions.

Finally we would like to compare between the micro-rheometer based on birefringence \cite{haward2012optimized}, and the proposed rheometer. Both concern extensional rheology, can be scaled down and are applicable for inline use, but as listed in table \ref{tab} there are also a number of significant differences.

\begin{table}
\begin{tabular}{l l l}
Property & Proposed & Birefringence \cite{haward2012optimized} \\ \hline
Dilute fluids & $ \times $ & \checkmark \\ %No & Yes \\
Concentrated fluids & \checkmark & $\times$ \\ %Yes & No \\
Simple setup & \checkmark & $\times$ \\ %Yes & No \\
Gives a curve & $\times$ & \checkmark \\ %No & Yes \\
Bistability is a & feature & limitation
\end{tabular}
\caption{Two extensional micro-rheometers are compared, the proposed and the one based on birefringence \cite{haward2012optimized}.}\label{tab}
\end{table}

\section{Conclusion}
We have described the background for generating catastrophes in bistable viscoelastic systems on a general level and presented simulation results in which the inlet flow rates in the cross-slot geometry is used as parameters to generate the catastrophe. Finally, we have discussed pros and cons for the suggested rheometer as well as provided guidelines for experimental verification. 



%% References


%\bibliographystyle{elsarticle-num}
%\bibliography{_Main}



\providecommand{\noopsort}[1]{}\providecommand{\singleletter}[1]{#1}%

\end{document}